\documentstyle[preprint,aps, eqsecnum, amssymb, epsfig]{revtex}
\begin{document}
\tightenlines
\draft

\title{Conserving and gap-less model of the weakly interacting Bose gas}
\author{Martin Fliesser,$^1$ J\"urgen Reidl,$^1$ P\'eter Sz\'epfalusy,$^{1,2}$
 and Robert Graham $^1$}
\address{$^1${\it Fachbereich Physik, Universit\"at Gesamthochschule
Essen,
45117 Essen, Germany}\\
$^2${\it Department of Physics of Complex Systems,
E\"otv\"os University, P\'azm\'any P\'eter s\'et\'any 1/A,
H-1117 Budapest, Hungary,\\
and
Research Institute for Solid State Physics and Optics,
P.O. Box 49, H-1525 Budapest, Hungary}}
\date{\today}
\maketitle

\begin{abstract}
The dielectric formalism is used to set up an approximate
description of a spatially homogeneous weakly interacting Bose gas in the
collision-less regime, which is both conserving and gap-less, and has
coinciding
poles of the single-particle Green's function and the density autocorrelation
function in the Bose-condensed regime. The approximation takes into account
the direct and the exchange interaction in a consistent way.
 The fulfillment of the generalized Ward identities related to the
conservation of particle-number and the breaking of the gauge-symmetry is
demonstrated. The dynamics at long wavelengths is considered in
detail below and above the phase-transition,  numerically and in certain
limits also in analytical approximations. The explicit form of the density
autocorrelation function and the Green's function is exhibited and discussed.

\end{abstract}


\vskip2pc
\narrowtext

\section{Introduction}
The observation of Bose-Einstein condensation \cite{Obs} in trapped systems
has given rise to an extensive experimental \cite{ExpRev,ExpRev2}
and theoretical
\cite{TheoRev}
study of weakly interacting dilute Bose-condensed gases.
Recent measurements of the elementary excitations \cite{Jila,Me96}
permit the detailed comparison with different theoretical approaches.
Due to the existence of a number of essential experimental parameters in the
 trapped systems,
e.g. the trap frequencies, scattering length, size of the condensate and
temperature,
the validity of different approximations depending on these conditions can be
examined.
Most of the theoretical approaches go back to the 1960s,
and were originally
derived for spatially homogeneous systems
in view of their application to superfluid helium.
For their application to the modern Bose-Einstein condensates
in traps many of these approaches are now being
extended to  inhomogeneous systems
and new approaches are also being developed.
In the present paper we wish to explore  such a new approximation scheme.
It is of interest for the modern Bose-Einstein condensates, because
measurements of quasi-homogeneous properties can also be made there \cite{Andrews}.

From the theoretical point of view taking the homogeneous limit
implies a great simplification.
Due to the translation symmetry complicated integral
equations reduce to
 algebraic equations;
essential  requirements
like a gap-less single-particle  spectrum
(which applies only in the homogeneous system)
and
the particle number conservation can be expressed in the form
of algebraic identities ({\sl Hugenholtz Pines theorem \cite{HoMa}, generalized
Ward identities \cite{wong}, compressibility sum-rule \cite{Forster}, etc.}).
In this form they provide important tests
for the consistency of approximations
obtained by selecting certain subsets of Feynman diagrams.
Having defined an approximation or `model' in this way analytical or
numerical solution is possible in the spatially homogeneous case without the
need for further approximation and all branches of the collective modes
- damped, strongly damped and even over-damped (purely damped) -
can be exhibited. This is a great advantage over
the corresponding calculations for the spatially inhomogeneous system,
 where further approximations are unavoidable
and a similarly complete
picture usually cannot be attained.

The  purpose of this paper is to examine the homogeneous limit of
an approximation previously derived for and applied to
the inhomogeneous case \cite{Rd99b}.
In contrast to the previous perturbative solution in
the case of a harmonic trap \cite{Rd99b}
we shall now avoid all further approximations, after setting up the
model by selecting the relevant class of Feynman diagrams, and
derive and study  all the excitation branches appearing in the
homogeneous model.
Furthermore, a number of exact theorems for the homogeneous system
will be checked for their validity within our model approach.

For example, in general the poles of the density autocorrelation function and
the single-particle Green's function must coincide in the Bose-condensed
region
(see e.g. \cite{Griffin}). Physically this is a consequence of the existence
of the order-parameter
describing the spontaneously broken gauge-symmetry.
In our approximation the required coincidence of the poles is a direct
consequence
of the derivation of the model within
the dielectric formalism,
which is designed to ensure just that.
In contrast to a previous model studied within this formalism (\cite{Sz74},
\cite{Griffin} and references therein)
and therefore sharing this crucial property,
the approximation or model studied here
(and in \cite{Rd99b} for the trapped system)
takes into account the contributions
of both the Hartree terms {\it and} the Fock (exchange) terms
 in the selected graphs.
Both types of terms are in fact of equal magnitude for the interaction via
$s$-wave scattering in the experimentally
realized dilute weakly interacting Bose condensates.

In  the next section we present
the basic diagrammatic building blocks
within the dielectric formalism as an approximation of the full system.
The basic relations of the dielectric formalism are then used
to
obtain the final density autocorrelation and Green's functions
determining
the excitation spectrum. We also state there the
{\it Generalized Ward identities}, which
are exact relations
ensuring the particle number conservation law in cases of spontaneously
broken symmetries.
We prove in  the appendix  that these identities continue to hold in  our model, which is a corner-stone of our model-building, because
it provides an important consistency check. A further consistency check is
the fulfillment of the compressibility sum-rule, which is demonstrated at the end of section II.
In Section \ref{sec:numerics} we present the numerical results for the
excitation branches at long wavelengths.
Besides the modes we studied previously for the inhomogeneous system
we find additional excitation branches
belonging to excitations of mainly the thermal density.
The behavior of these thermal branches will be studied in detail.
For weak interaction they can be accurately approximated by  simple analytic
expressions, which is done in Section \ref{analytic}.
For  stronger interaction only numerical results are obtainable.
They may suggest
 explanations
for the appearance of  the two different  branches measured at JILA \cite{Jin97}.
 Section \ref{conclusion} contains some conclusions and further remarks.

\section{THE MODEL}
\label{sec:form}
\subsection{The dielectric formalism}

Here we shall give some necessary background on the so-called `dielectric
formalism' at finite temperature, which was introduced  for analyzing interacting
Bose-condensed systems \cite{KS,MGW,Sz74,CG73,TG,Payne,Griffin}.
An excellent exposition of the formalism and a complete list of references is
given by Griffin \cite{Griffin}.
The presence of a Bose condensate breaks the
gauge-symmetry of
the system. As a consequence the density fluctuation spectra appear in
the one particle excitation spectra and vice versa.
In the dielectric formalism the corresponding
coupling mechanism is explicitly exhibited.

For sake of simplification we adopt
the natural unit $\hbar=1$ throughout the paper.
In the Bose-condensed phase the
usual decomposition of the Bose field operator
$\hat\Psi$ in a   condensate wave function
 $\Phi_0=\langle \hat\Psi (\bbox{r})\rangle$
and the  fluctuation operator
$\hat\Phi$ with vanishing expectation value
\begin{equation}
\hat\Psi= \Phi_0+ \hat\Phi
\end{equation}
is made.
$\langle \ldots \rangle$ denotes the thermal averaging
\begin{equation}
\langle \hat{A} \rangle = {\mbox{Tr}\, \hat{A}
e^{-\beta(\hat{H}-\mu\hat{N})}\over \mbox{Tr}\,
e^{-\beta(\hat{H}-\mu\hat{N})}}.
\end{equation}
The condensate wave function can in general be complex, but in the absence of vortices it can be chosen as real, without restriction of generality, and we shall consider this case in the following, for simplicity.
The value of the chemical potential $\mu$ is fixed by the
average number of particles in the system and is connected with the density of particles in the condensate $|\Phi_0|^2$ by
requiring $\langle\hat\Phi\rangle=0.$
Imposing $\Phi_0\ne 0$ we have the additional
appearance of anomalous Green's functions and it is useful to introduce the matrix Green's function
\begin{equation}
G_{\alpha\beta}(\bbox{r},\tau;\bbox{r}',\tau')=-
\langle T_\tau\left[ \hat\Phi_\alpha(\bbox{r},\tau)
\hat\Phi_\beta^\dagger(\bbox{r}',\tau')\right]\rangle
\label{eq:gfdef}
\end{equation}
with field operators in Matsubara representation $\hat\Phi_1
(\bbox{r},\tau) \equiv \hat\Phi(\bbox{r},\tau)$ and
$\hat\Phi_2 (\bbox{r},\tau) \equiv
\hat\Phi^\dagger (\bbox{r},\tau)$.
In addition to these one particle correlation functions
we need  the density autocorrelation function:
\begin{equation}
\chi_{\rm nn} (\bbox{r},\tau;\bbox{r}',\tau') =-
\langle T_\tau \left[
\tilde{n}(\bbox{r},\tau)\tilde{n}(\bbox{r}',\tau')\right]\rangle.
\label{eq:chidef}
\end{equation}
Here $\tilde{n}= \hat{n}-\langle \hat{n}\rangle$ denotes the density deviation
operator.
The exact Green's function $G_{\alpha\beta}$
can be obtained from Dyson's equation, i.e. by summing up a geometric series
in terms of the Green's function $G^0_{\alpha\beta}$ of the noninteracting
system and the
 self-energy $\Sigma_{\alpha\beta}$:
\begin{equation}
G_{\alpha\beta}=G^0_{\alpha\beta}+G^0_{\alpha\gamma}
\Sigma_{\gamma\delta}
G_{\delta\beta}.\label{Ge}
\end{equation}
The self-energy insertions are by definition {\sl irreducible} i.e.
{\sl cannot be split  into two parts by
cutting a single particle line}.
Different model approximations can be classified by the
different kinds of
 interaction processes  included in $\Sigma_{\alpha\beta}$.

Beyond the definition of irreducible quantities
 it is equally  useful to
consider the {\it proper} contributions. Diagrams are called proper if they
 {\it cannot be split into two parts, each connected to an external vertex or line, by cutting a single
interaction line}. They will be denoted
by an additional {\it tilde}.
The total density autocorrelation function
${\chi_{\rm nn}}$ can be easily recovered
from the proper contributions $\tilde{\chi}_{\rm nn}$
by just summing
up a geometric series:
\begin{equation}
\label{chiser2}
\chi_{\rm nn}(\bbox{q},\omega)=\frac{\tilde{\chi}_{\rm nn}(\bbox{q},\omega)}
{1-g\tilde{\chi}_{\rm nn}(\bbox{q},\omega)
}.
\end{equation}
Here $\omega$ on the one hand denotes the Matsubara-frequency \cite{FW}
$\omega=i\omega_{\rm n}=i(2n\pi/\beta
)$.
On the other hand it is  understood that the analytic continuation in $\omega$
from the positive imaginary axis
yields the retarded correlation
functions, and therefore, if this is our purpose, $\omega$ will in the
following always be assumed to lie in the upper half of the complex plane.
$g$ denotes the 2-particle T-matrix.
In the whole temperature-domain of relevance here it can be taken as independent of $\bf{q}, \omega$ and be expressed by the s-wave scattering length $a$ as
\begin{equation}
g=\frac{4\pi a}{m}
\end{equation}
The importance of the use of the T-matrix in place of the bare interaction was first stressed in this context in Beliaev's work \cite{Bel}. (See \cite{CG73} for a discussion of ladder diagrams within the dielectric formalism at finite temperature.) As is well known
since then the use of the 2-particle T-matrix in place of the bare interaction implies that a sum is implicitly performed over the ladder-diagrams of the two-particle interaction, and care has to be taken not to sum up the same class of diagrams again later-on, see e.g. \cite{ShiGr}.

In the dielectric formalism
the approximations for $\Sigma_{\alpha\beta}$ and
for $\tilde{\chi}_{\rm nn}$ are related to each other via irreducible
vertex functions $\Lambda_\alpha$ describing single
processes of excitations out of the condensate and of relaxations into
the condensate where the energy and momentum transfer is described
by an additional out-going interaction line.
Unlike  in the usual Green's function approach, where approximations are
defined by the diagrams kept in the self-energies and $\tilde{\chi}_{\rm nn}$
as  basic building blocks, in the
dielectric formalism an approximation (or `model')
is defined by introducing
building blocks  for the proper and irreducible parts of the three quantities
$\Sigma_{\alpha\beta}$, $\chi_{\rm nn}$ and $\Lambda_\alpha$.
 In the following we call
{\it irreducible and proper} quantities to be {\it regular} and denote them
with an $r$ in brackets.

Thus we start  with specifying
the building blocks ${\chi}_{\rm nn}^{(r)}$,
${\Sigma}_{\alpha\beta}^{(r)}$ and ${\Lambda}_\alpha^{(r)}$ and
expressing all necessary quantities
in terms of these three basic quantities, see e.g. \cite{Griffin}.
We begin with the decomposition of the proper quantities
$\tilde{G}_{\alpha\beta}$
and $\tilde{\chi}_{\rm nn}$.
Since a Green's function  is proper if and only if all the
self-energy insertions are proper we get
\begin{equation}
\tilde{G}_{\alpha\beta}^{-1}(\bbox{q},\omega)
=(G_{\alpha\beta}^0)^{-1}(\bbox{q},\omega)-
{\Sigma}_{\alpha\beta}^{(r)}(\bbox{q},\omega)
\end{equation}
with $(G_{\alpha\beta}^0)^{-1}(\bbox{q},\omega)=\delta_{\alpha\beta}\left[
\alpha\omega-(\frac{
\bbox{q}^2}{2m}-\mu
)\right]$
where $m$ is the mass of the atoms and we use the abbreviation $\alpha=1$
or -1
if the index $\alpha$ takes the values 1 or 2, respectively.
Next we consider
the contributions to $\tilde{\chi}_{\rm nn}$, which can be written
in the form:
\begin{equation}
\label{tchiz}
\tilde{\chi}_{\rm nn}=\chi_{\rm nn}^{(r)}+{\Lambda}_\alpha^{(r)}
\tilde{G}_{\alpha\beta}{\Lambda}_\beta^{(r)}.
\end{equation}
Here we have both,
irreducible contributions ${\chi}_{\rm nn}^{(r)}$ and
contributions which contain at least one proper Green's function
$\tilde G_{\alpha \beta}$.

The  expressions for ${\chi}_{\rm nn}$ and
$\Sigma_{\alpha\beta}$
are obtained by summing up the corresponding
geometric series:
The result for $\chi_{\rm nn}$ was already given in (\ref{chiser2}), and we can
immediately turn to
 the elements of $\Sigma_{\alpha\beta}$, which we decompose
into their proper and their improper parts as follows: \begin{eqnarray}
\label{Sigmadef}
\Sigma_{\alpha\beta}(\bbox{q},\omega)&=&
{\Sigma}_{\alpha\beta}^{(r)}(\bbox{q},\omega)
+{\Lambda}_\alpha^{(r)}(\bbox{q},\omega)\frac{g}
{1-g\chi_{\rm nn}^{(r)}(\bbox{q},\omega)
}
{\Lambda}_\beta^{(r)}(\bbox{q},\omega).
\end{eqnarray}

$\langle\hat\Phi\rangle=0$ implies that $\Sigma_1^{(r)}(0,0)=0$, where $\Sigma_1^{(r)}(\bbox{q},\omega)$ is the sum of the regular
 diagrams with one external line (tadpole-diagrams).

In addition the conservation laws and corresponding sum rules must be
woven into the formalism by ensuring that the approximate expressions
for the regular parts still satisfy the exact generalized Ward identities.
The identities
we want to consider  in the following are a consequence of particle
number conservation which becomes nontrivial because of the broken gauge-symmetry.
Within the dielectric formalism the corresponding Ward-identities
were first introduced by
Wong and Gould \cite{wong} at temperature $T=0$ and later developed further for finite temperature (see  e.g. \cite{TG,Griffin} and further references given there) and used as exact relations between the building-blocks of the dielectric formalism to be satisfied
by any consistent  approximation. They can be obtained by
inserting the continuity equation
\begin{equation}
\label{continuity}
\frac{\partial\hat{n}(\bbox{q},\tau)}{\partial\tau}=
-\frac{\bbox{q}}{m}\cdot\hat{\bbox{J}}(\bbox{q},\tau)
\end{equation}
into the expressions for various time-ordered correlation functions \cite{wong},\cite{Griffin}
and read:
\begin{eqnarray}
\label{ward1}
\omega{\Lambda}^{(r)}_\alpha(\bbox{q},\omega)&=&\frac{|\bbox{q}|}{m}
{\Lambda}^{l\,(r)}_\alpha(\bbox{q},\omega)+\sqrt{n_c}\left[
\omega-\alpha(\frac{\bbox{q}^2}{2m}-\mu)\right]
-\sqrt{n_c}\left[{\Sigma}^{(r)}_{1,\alpha}(\bbox{q},\omega)
-{\Sigma}^{(r)}_{2,\alpha}(\bbox{q},\omega)\right],\\
\label{ward2}
\omega{\chi}^{(r)}_{\rm nn}(\bbox{q},\omega)&=&\frac{|\bbox{q}|}{m}
{\chi}^{l\,(r)}_{Jn}(\bbox{q},\omega)-
\sqrt{n_c}\left[
{\Lambda}^{(r)}_1(\bbox{q},\omega)-{\Lambda}_2^{(r)}
(\bbox{q},\omega)\right],\\
\label{ward3}
\omega{\chi}^{l\,(r)}_{Jn}(\bbox{q},\omega)&=&\frac{|\bbox{q}|}{m}
[{\chi}^{l\,(r)}_{JJ}(\bbox{q},\omega)+
m(n_c+n_{\rm th})]-
\sqrt{n_c}\left[
{\Lambda}^{l\,(r)}_1(\bbox{q},\omega)
-{\Lambda}^{l\,(r)}_2(\bbox{q},\omega)\right].
\end{eqnarray}
The demonstration that eqs(\ref{ward1}-\ref{ward3}) are satisfied by our model approximation is crucial for ensuring its consistency and given in the appendix.

A further important requirement for the consistency of any approximation
is the compressibility sum-rule
\cite{Forster}
\begin{eqnarray}
\lim\limits_{|\bbox{k}|\to 0} \chi_{\rm nn}({\bbox{k}},0)=-\frac{\partial
n_{\rm tot}}
{\partial \mu}{\displaystyle |_T}.
\label{csr1}
\end{eqnarray}
To these exact requirements should in our opinion be added a further one, 
which is a consequence of Galilei-invariance and will be discussed 
in the final section.

\subsection{The Model}

We shall  now introduce and motivate the particular approximation (or 'model') within the framework of the dielectric formalism we wish to study in this paper. Our aim here is to give a self-consistent  Hartree-Fock theory of the Bose-gas,
not a perturbation theory in the coupling constant to any given order.
This aim seems legitimate in view of the success of the Hartree-Fock approximation in many-body physics in general, its limitations as a mean-field theory notwithstanding.

Let us first consider the Bose-gas in the non-condensed phase. Previously this system was studied in the framework of the dielectric formalism within the Hartree-approximation \cite{Sz74}, which corresponds to the self-consistent
theory summing all the bubble-diagrams, but neglecting exchange-contributions in the thermal cloud.. Here we would like to go a step further and study
the self-consistent Hartree-Fock theory  summing the bubble-diagrams {\it and} the exchange-bubble diagrams. For the uncondensed phase this approximation is, of course, standard, both for the single-particle
Green's function, where it corresponds to the standard Hartree-Fock approximation, and for the density response-function, where the summing of the exchange-bubbles is discussed e.g. in \cite{FW}.

{}The dynamics of the thermal cloud is described by the Hartree-Fock Green's
function, which takes the form
\begin{equation}
   G^{HF}_{\alpha\beta}({\bbox{k}},\omega)={
\delta_{\alpha\beta}
\over\alpha
\omega-E^{HF}(k)}\,.
\label{GHF}
\end{equation}
with 
\begin{equation}
E^{HF}(k)=\frac{
{\bbox{k}}^2}{2m}
   +2g n_{tot}-\mu
\end{equation}

With these  Green's functions the lowest order contributions
to the regular density autocorrelation function $\chi_{nn}^{(r)}$ (see fig.\ref{fig:chila}) are given by the
so-called bubble-diagram
\begin{equation}
\chi_{\rm nn}^{0}(\bbox{q},\omega)=-{1 \over \beta
}
\sum_n {1 \over (2\pi
)^3}\int d^3{\bbox{p}}\,
G^{HF}_{11}({\bbox{p}},i\omega_{\rm n})\,
G^{HF}_{22}(\bbox{q}-\bbox{p},\omega-i\omega_{\rm n}).
\label{eq:chi0}
\end{equation}
For the sake of better readability we will use the abbreviations
$p\equiv({\bbox{p}},i\omega_{\rm n})$ representing the
momentum ${\bbox{p}}$ and the Matsubara frequency $i\omega_{\rm n}$.

In order to treat the exchange-contribution of the two-particle scattering in $\chi_{nn}$ on the same footing as the direct contribution we have to sum up the diagrams for $\chi_{nn}^{(r)}$ as in fig.\ref{fig:chila}, which define a simple geometric series. Since the
interaction $g$ is given here just by a constant the result can be written in terms of the expression (\ref{eq:chi0}) in the form 
\begin{equation}
\label{chirser1a}
\tilde\chi_{\rm nn}(q)=\frac{\chi_{\rm nn}^{0}(q)}{1-g
\chi_{\rm nn}^{0}(q)}.
\end{equation}
When this result for $\tilde\chi_{\rm nn}$ is substituted in eq.(\ref{chiser2})
it yields simply
\begin{equation}
\chi_{nn}(q)=\frac{\chi^0_{nn}(q)}{1-2g\chi^0_{nn}(q)},
\label{ch}
\end{equation}
i.e. a doubling of the effective coupling constant in the denominator. Concerning this treatment one may legitimately ask, why the Hartree-Fock Green's functions rather than the free-particle Green's functions are used for the internal propagators in the bubble and exchange-bubble diagrams. The reason is one of consistency. Without the use of Hartree-Fock propagators one could not achieve
 consistency between statics and dynamics,
which is incorporated in the compressibility sum-rule (\ref{csr1}) (nor,
as we may add, the consistency with
Galilei-invariance, as we discuss in the final section).

Let us now turn to the description of our approximation, or 'model',
also for the condensed phase. It is defined as the natural extension of the
Hartree-Fock model we use in the uncondensed phase. Thus, for the internal lines we use again the 
Hartree-Fock propagators. In this way we again achieve consistency with the compressibility sum-rule as will be shown later.  Furthermore, we again sum up the bubble and exchange-bubble diagrams for all proper quantities, which now also include the proper vertex
function $\Lambda^{(r)}$. Therefore we find
\begin{equation}
\label{chirser}
\chi_{\rm nn}^{(r)}(q)=\frac{\chi_{\rm nn}^{0}(q)}{1-g
\chi_{\rm nn}^{0}(q)},
\end{equation}
which is almost the same expression as found in the uncondensed phase, see also \cite{CG}.

For the  proper vertex function $\Lambda_\alpha^{(r)}(q)$
the zero-loop diagram is given by the
trivial vertex function $\Lambda^0_\alpha=\Phi_0$. By the same reasoning as for $\chi_{nn}^{(r)}$
we obtain a  further geometric series:
\begin{equation}
\label{tlambda}
\Lambda^{(r)}(q)=\frac{\Phi_0}{1-g
\chi_{\rm nn}^0(q)},
\end{equation}
illustrated in Fig.\ref{fig:vertexla}.

\begin{figure}
\begin{center}
\epsfig{file=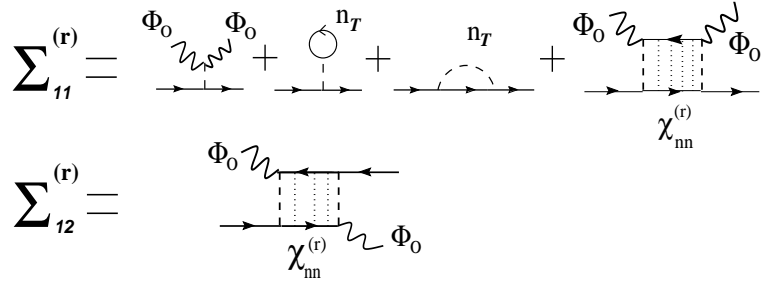, height=10cm, angle=270}
\end{center}
\caption{Contributions to   $\Sigma^{(r)}_{\alpha\beta}$.}
\label{fig:selfla}
\end{figure}

The regular self-energy contributions given by
$\Sigma^{(r)}_{\alpha\beta}(q)=\Sigma^{0\,(r)}_{\alpha\beta}(q)
+\Sigma^{1\,(r)}_{\alpha\beta}(q)$ are presented in Fig.\ref{fig:selfla},
 where
\begin{equation}
\label{tsigma0}
{\Sigma}^{0\,(r)}_{\alpha \beta}=\left(
g|\Phi_0|^2
+2gn_{\rm th}\right)
\left(\begin{array}{rl}
1&0\\
0&1
\end{array}\right)_{\alpha \beta}.
\end{equation}
\begin{equation}
\label{tsigma1}
\Sigma^{1\,(r)}_{\alpha\beta}(q)=\left(\frac{g\Phi_0^2}{1-g\chi_{nn}^0(q)}
-g\Phi_0^2\right)
\left(\begin{array}{rl}
1&1\\
1&1
\end{array}\right)_{\alpha \beta}\,,
\end{equation}

 We now simply insert the expressions
(\ref{tsigma0}), (\ref{tsigma1}), (\ref{tlambda}) and (\ref{chirser})
in Eq. (\ref{Sigmadef}),
and obtain straightforwardly the expressions for $\Sigma_{\alpha\beta}(q)$
of our approximation, which are not written out here to conserve space.
Using our results for 
$\Sigma_{11}$ and $\Sigma_{12}$ it can be easily checked that the
Hugenholtz-Pines theorem \cite{FW} $\mu=\Sigma_{11}(0)-\Sigma_{12}(0)$
is satisfied.

\begin{figure}
\begin{center}
\epsfig{file=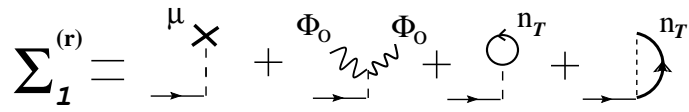, height=9cm, angle=90}
\end{center}
\caption{Contributions to   $\Sigma^{(r)}_{1}$.}
\label{fig:tadp}
\end{figure}

The tadpole diagrams are given in fig.\ref{fig:tadp}. The condition $\Sigma_1^{(r)}=0$  leads to a relation which can be solved for the chemical potential 
\begin{equation}
\mu=gn_c+2gn_{\rm th}+\mu^{(0)},
\label{mu}
\end{equation}
where we added the chemical potential $\mu^{(0)}$ of the ideal Bose-gas,
which vanishes in the Bose-condensed phase, in order to make the expression valid also in the uncondensed phase, and
where we define
\begin{equation}
n_{\rm th}=-\sum_n\int \frac{d^3p}{(2\pi)^3}G^{HF}({\bf p},i\omega_n).
\end{equation}
That $n_{\rm th}$ is consistently interpreted as the thermal density of the particles in our approximation will be shown by checking the Ward-identities and the compressibility sum-rule.

One might raise the question why in the diagrams for $\Sigma^{(r)}_1$ and the diagonal elements $\Sigma^{(r)}_{\alpha\alpha}$ exchange-ladder terms with two condensate lines at the same interaction line are not also included.
The reason is that because of our use of Hartree-Fock propagators in the internal lines these exchange-diagrams are already contained in the first-order proper exchange-diagram.
As a result the chemical potential $\mu$ in eq.(\ref{mu}) is simply given by the lowest order terms, but of course these must be evaluated with the Hartree-Fock propagators.

Using the building-blocks we have specified it is now straight-forward to
obtain the explicit expressions for the Green's function,
namely
\begin{eqnarray}
G_{11}({\bbox{k}},\omega)&&=\frac{(\omega+\frac{
\bbox{k}^2}{2m}-\mu^{(0)}
)(1-2g
\chi^0_{\rm nn}(\bbox{
k},\omega))+g
n_c(1+2g
\chi^0_{\rm nn}(\bbox{
k},\omega))}{\Delta({\bbox{k}},\omega)}\nonumber\\
G_{12}(\bbox{
k},\omega)&&=-\frac{g
n_c(1+2g
\chi^0_{\rm nn}(\bbox{
k},\omega))}{\Delta({\bbox{k}},\omega)}\label{Green}
\end{eqnarray}
with the denominator
\begin{equation}
\Delta({\bbox{k}},\omega)=(\omega^2-(\frac{
\bbox{k}^2}{2m}-\mu^{(0)}
)^2)(1-2g
\chi^0_{\rm nn}(\bbox{
k},\omega))-2g
n_c\frac{
\bbox{k}^2}{2m}(1+2g
\chi^0_{\rm nn}({\bbox{k}},\omega)).
\end{equation}
The density autocorrelation function can be given similarly as
\begin{equation}
\chi_{\rm nn}({\bbox{k}},\omega)=\frac{(\omega^2-(\frac{
\bbox{k}^2}{2m}-\mu^{(0)})^2)
\chi^0_{\rm nn}(\bbox{
k},\omega)+2
n_c\frac{
\bbox{k}^2}{2m}
(1+g
\chi^0_{\rm nn}({\bbox{k}},\omega))}{\Delta(\bbox{
k},\omega)}\,.\label{auto}
\end{equation}
The poles of the Green's function and
the density autocorrelation function are  given by $\Delta(\bbox{
k},\omega)=0$. In the uncondensed phase we can put $n_c=0$ and the result (\ref{auto})
reduces to eq.(\ref{ch})
with the poles given by $g
\chi^0_{\rm nn}({\bbox{k}},\omega)=1/2$.
These dispersion-relations will be analyzed in section
\ref{sec:numerics}.

It is manifest from our explicit expressions
that the poles of the Green's function in the Bose-condensed phase are indeed the same
as those of the density autocorrelation function.
On the other hand, because of $n_c=0$  the poles of $\chi_{\rm nn}({\bbox{k}},\omega)$ and
$G_{11}({\bbox{k}},\omega)$ become different in the uncondensed phase: the factor $(\omega+\frac{
\bbox{k}^2}{2m}-\mu^{(0)}
)(1-2g
\chi^0_{\rm nn}(\bbox{
k},\omega))$
then cancels from the numerator and denominator of $G_{11}({\bbox{k}},\omega)$
and the remaining single-particle poles at $\omega=(
\bbox{k}^2/2m)-\mu^{(0)}
$ are those of the free Bose-gas.

It is important to remark here, that summing up the bubble and exchange-bubble diagrams, after a suitable rearrangement of the resulting expressions, has not lead us to terms of higher than the first loop order in the numerator and denominator, both in the condensed and the uncondensed phase, if note is made of the fact that the quantity $gn_c$
is of zeroth loop-order in the condensed phase.
However, our treatment is not a perturbation theory in the coupling g, i.e. loop-order is not the same as order in the coupling $g$, because
a self-consistently determined propagator, namely the Hartree-Fock
propagator, is used in the bubble-diagram defining $\chi_{nn}^0$, which depends itself on $g$. Rather, we claim that the theory given here is the self-consistent Hartree-Fock theory also in the condensed phase.
This claim is based on the fact that the theory satisfies all consistency checks, namely the identity of the poles of the Green's- and density response-function, the Ward-identities, which also contain the f-sum rule,  the compressibility sum-rule, and of course the Hugenholtz-Pines theorem. In addition it satisfies a consistency check derived from Galilei-invariance, as discussed in the final section. It should be noted that, apart from the Hugenholtz-Pines theorem,
none of these consistency checks are satisfied in the usual Popov approximation. (But it is  interesting to remark that they can 
also be shown to be satisfied in the simpler Hartree-model \cite{Sz74,Payne} 
within the dielectric formalism, for which our treatment  therefore provides the logical consistent extension which includes exchange).

Surprisingly for us the comparison of our result (\ref{auto}) for the
density autocorrelation function with corresponding results given
by Minguzzi and Tosi \cite{Min} yields complete agreement. 
This is  nontrivial, since the theories
are formulated in quite different ways and arrive at the result for the
density autocorrelation function in very different manners and after considerable rearrangements. It is
interesting to learn from this agreement that the theory
formulated by Minguzzi and Tosi \cite{Min} has a foundation within the
dielectric formalism and permits a
description in terms of the Feynman diagrams from which we derived our results.
This is a strong hint that there is just one consistent Hartree-Fock theory 
not only in the uncondensed but also in the condensed phase.

An obvious advantage of the dielectric formalism  over the
linear response theory used in \cite{Min} is the fact
that besides the density response-function the Green's function 
 is also obtained explicitly as displayed above, which
would be outside the scope of the
theory given in \cite{Min}.
With the Green's function our theory gives explicit expressions for  autocorrelation
functions of amplitude and phase of the order-parameter.
The former equals $-G_+$, the latter $-G_-$, with
\begin{eqnarray}
G_+({\bbox{k}},\omega)&&=G_{11}({\bbox{k}},\omega)+G_{12}({\bbox{k}},\omega)\nonumber\\
&&=\frac{(\omega+\frac{
\bbox{k}^2}{2m}-\mu^{(0)}
)(1-2g
\chi^0_{\rm nn}(\bbox{
k},\omega))}{\Delta({\bbox{k}},\omega)}\\
G_-({\bbox{k}},\omega)&&=G_{11}({\bbox{k}},\omega)-G_{12}({\bbox{k}},\omega)\nonumber\\
&&=\frac{(\omega+\frac{
\bbox{k}^2}{2m}-\mu^{(0)}
)(1-2g
\chi^0_{\rm nn}(\bbox{
k},\omega))+2g
n_c(1+2g
\chi^0_{\rm nn}(\bbox{
k},\omega))}{\Delta({\bbox{k}},\omega)}.
\end{eqnarray}

We still need to evaluate
the thermal density $n_{\rm th}$, the chemical potential
$\mu$ and $n_c$.\\
By integrating the Bose factor
\begin{equation}
   f_0(E^{HF}(k))
   = \frac{1}{\exp{[\beta E^{HF}(k)]}-1}
\label{bose.fact}
\end{equation}
over the momenta
we get $n_{\rm th}=g_{3/2}(z)/\lambda_{\rm th}^3\,$,
where $g_{3/2}(z)$ is the Bose function \cite{FW}, $z$ is the fugacity,
defined in terms of an effective chemical potential
$\bar\mu=\mu-2gn_{\rm tot}$ via
$z=\exp(\beta\bar\mu)$, and
$\lambda_{\rm th}=\sqrt{2\pi
/mk_BT}$
denotes the thermal wavelength. In the uncondensed region the effective chemical potential
$\bar\mu$ coincides with the chemical potential $\mu^{(0)}$ of the ideal
Bose-gas. For the condensed phase,
using $\mu=g\,n_c + 2g\,n_{\rm th}$,
the effective chemical potential is
$\bar\mu=\mu-2g\,(n_{\rm th}+n_c)
=-gn_c$ and  $z=e^{-\beta gn_c}$.
For a given total density
$n_{\rm tot}=n_c+n_{\rm th}$
the equation of state in the condensed phase is therefore given by the
implicit equation, which is equivalent to the equation of state
derived first by Huang et al. \cite{Huang},
\begin{equation}
   n_{\rm tot}
   =n_c +\frac{1}{\lambda_{\rm th}^3}\,g_{3/2}(e^{-\beta g n_c})\,.
\label{ncond.impl}
\end{equation}

It can be seen from this equation that
within our approximation for $G^{HF}$,
the phase transition
is no longer continuous\footnote {A discontinuity  is also found in the
familiar  Popov-approximation \cite{Popov,Griffpop}, see e.g.
\cite{Leggett,Siggia,ShiGr,Pomeau}}.: E.g. for $T=T_c$, where $T_c$ is defined as the critical temperature of the free
Bose-gas for the same particle-density by $n_{tot}=g_{3/2}(1)/\lambda^3_{\rm th}(T_c)$, there are two solutions $n_c=0$
and $n_c=\Delta n_c$, with
\begin{equation}
   \Delta n_c(T_c)\, (\lambda_{\rm th}^c)^{3}
 = \frac{4\pi \frac{g (\lambda_{\rm th}^c)^{-3}}{k_BT_c}}{(1-\zeta(1/2)\frac{g (\lambda_{\rm th}^c)^{-3}}{k_BT_c})^2}
   + O((\frac{g (\lambda_{\rm th}^c)^{-3}}{k_BT_c})^{5/2}).
\label{ncond.disc}
\end{equation}
$\lambda_{\rm th}^c$ denotes the thermal wavelength at the
critical temperature.
Even for comparably weak interaction $g n_{\rm tot}/k_BT_c =0.1$
this jump already amounts to $16\%$ of the total density,
so it cannot be neglected by any means. The coexisting values of the total density for given values of the chemical potential in the two-phase region
have to be determined from the Maxwell-construction and will be discussed below.

Let us first consider the correlation functions in the static limit, i.e. at
$\omega=0$, in more detail. We find with or without condensate
\begin{equation}
G_{-}({\bbox{k}},0)=-\frac{1}{\frac{
\bbox{k}^2}{2m}-\mu^{(0)}}
\,.
\end{equation}
With a condensate $\mu^{(0)}=0$ and this correlation function displays the
usual infrared singularity $\sim \bbox{k}^{-2}$ associated to the
spontaneously broken gauge-symmetry.
For the amplitude autocorrelation function at $\omega=0$ we obtain in the uncondensed and the condensed phase,
respectively, and displaying also the asymptotics for $|\bbox{k}|\to 0$,
\begin{eqnarray}
G_{+}({\bbox{k}},0)&&=-\frac{1}{\frac{
\bbox{k}^2}{2m}-\mu^{(0)}
}\rightarrow\frac{1
}{\mu^{(0)}}\qquad  T>T_c\nonumber\\
\label{cor}\\
G_{+}({\bbox{k}},0)&&=-\frac{1
-2g\chi^0_{\rm nn}(\bbox{
k},0)}{\frac{
\bbox{k}^2}{2m}(1-2g
\chi^0_{\rm nn}(\bbox{
k},0))+2gn_c(1+2g
\chi^0_{\rm nn}(\bbox{
k},0))}\rightarrow\frac{-1}{2\Sigma_{12}({\bbox{k}},0)}\qquad  T<T_c\nonumber
\end{eqnarray}
with
\begin{equation}\frac{1}{\Sigma_{12}(\bbox{
k},0)}=\frac{1-2g
\chi^0_{\rm nn}({\bbox{k}},0)}
{gn_c(1+2g
\chi^0_{\rm nn}({\bbox{k}},0))}\,.
\end{equation}

\begin{figure}
\begin{center}
\epsfig{file=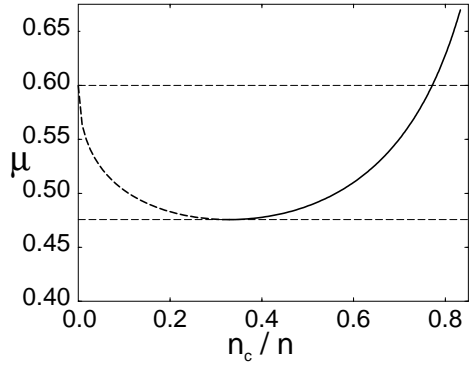, height=7cm, angle=270}
\end{center}
\caption{The chemical potential $\mu$ of the condensed phase
in units of $k_BT$ as a function of
the condensate fraction  $n_c/n$ for fixed temperature. The plot is made
for the coupling strength $g=0.3\lambda_{\rm th}^3k_B T /\zeta(3/2)$.
The horizontal long-dashed
lines represent the borders of multi-stability, the upper one given
by $\mu^{\rm (max)}=2g\lambda_{\rm th}^{-3}\zeta(3/2)$ the lower one
$\mu^{\rm (min)}$ by
the instability condition $g\chi(0,0)=-1/2$.
The stable part of the curve is plotted as
a solid line changing to a long-dashed line in the region of instability.}
\label{fig:muvnc}
\end{figure}

We examine the stability of the model  
below
$T_c$. As can be seen from the results for $G_{+}({\bbox{k}},0)$
an instability may occur if for small $|{\bbox{k}}|\to 0$ we satisfy
$g
\chi^0_{\rm nn}({\bbox{k}},0)=-1/2$.In order to see where  this condition for instability is satisfied
we evaluate $\chi^0_{\rm nn}(0,0)=
\lim\limits_{|\bbox{k}|\to 0}
\chi^0_{\rm nn}({\bbox{k}},0)$ from Eq. (\ref{eq:chi0}), finding
\begin{equation}
g
\chi_{\rm nn}^0(0,0)
=-\frac{g\beta}{\lambda_{\rm th}^3}g_{1/2}(e^{-\beta gn_c}).
\label{2.38}
\end{equation}
This is a negative quantity, which, for sufficiently low temperature, lies in
the interval $[-1/2, 0]$, but which moves monotonously
in the direction of the instability at -1/2 with decreasing condensate density
for fixed temperature.
The instability condition $g
\chi_{\rm nn}^0(0,0)=-1/2$ is equivalent to the
condition that for fixed temperature
the chemical potential $\mu$ exhibits a minimum $\mu^{\rm (min)}$
as a function
of $n_c$.
In Fig. \ref{fig:muvnc} we plot this functional behavior.
The fact that the instability condition, which we have obtained here from the dynamics described
by $G_{\alpha\beta}$, fits completely with the form of the equation of state
(\ref{ncond.impl}) underscores the consistency between statics and dynamics
in our model.

\begin{figure}
\begin{center}
\epsfig{file=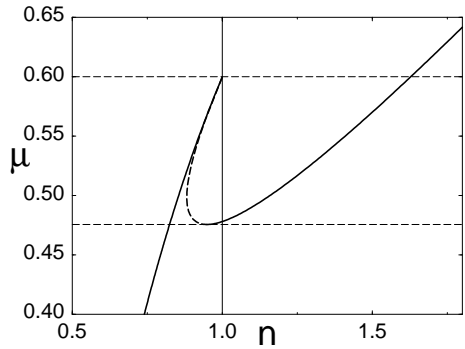, height=7cm, angle=270}
\end{center}
\caption{The chemical potential $\mu$ of the
condensed and the uncondensed Bose gas in units of $k_B T$ as
a function of the particle  density $n$
in units of $\zeta(3/2)\lambda_{\rm th}^{-3}$ for the same value of the
coupling constant $g$ as in Fig. \ref{fig:muvnc}.
The long-dashed curve represents
the unstable part of the branch of the condensed phase.}
\label{fig:muvn}
\end{figure}

In Fig. \ref{fig:muvn} we plot the
branch corresponding to the chemical potential of the
uncondensed Bose gas in Hartree-Fock approximation together with the branch for the
condensed system. We only get solutions for the uncondensed Bose
gas if $\bar{\mu}=\mu^{(0)}
\le0$ or equivalently $\mu\le \mu^{\rm (max)}
=2g \lambda_{\rm th}^{-3}\zeta(3/2)$.
Combining these results  due to the first order transition we find a
region of multi-stability  between the condensed and the uncondensed Bose
gas for chemical potential values $\mu^{\rm (min)}\le\mu\le\mu^{\rm (max)}$.
In this region there are actually two possible
non-vanishing values for the condensate density $n_c\ne 0 $
for given temperature and chemical potential.
But due to  the stability condition
$g\chi^0_{\rm nn}(0,0)\geq -1/2$
only one of these solutions
proves to be stable. The coexisting values of $n_{tot}$ at a given temperature are
fixed by the Maxwell-construction for the $\mu(n_{\rm tot})$-curve.
The resulting
value $\mu_{\rm coex.}$ of the chemical potential then also determines the
value of $n_c$ coexisting with $n_c=0$ from Fig. \ref{fig:muvnc}.

The equation of state as obtained from Eq. (\ref{ncond.impl})
and plotted in Fig.\ref{fig:muvn} is fully consistent with the
density-density correlation function (\ref{auto}):  the compressibility sum-rule (\ref{csr1}) makes this connection between
statics and dynamics. With our present result for $\chi_{nn}({\bf k},0)$ it takes the form
\begin{eqnarray}
\frac{\partial n_{\rm tot}}
{\partial \mu}{\displaystyle |_T}=\frac{1}{g}\,
\frac{1+g\chi^0_{\rm nn}(0,0)}{1+2g\chi^0_{\rm nn}(0,0)}
\label{csr2}
\end{eqnarray}
That the derivative $(\partial n_{tot}/\partial\mu)_T$ can indeed be written in this form can be checked by a short calculation using the identity
$\frac{\partial}{\partial \bar{\mu}}g_{3/2}
(e^{\bar{\mu}\beta})=-\beta g_{1/2}
(e^{\bar{\mu}\beta})$ and eqs.(\ref{ncond.impl},\ref{2.38}).
Thus, from the right-hand side of equation (\ref{csr2}), 
which is obtained from the dynamics,
a static property on its left-hand side, namely the equation of
state, can be re-derived by integration with respect to $\mu$.

The slope $\frac{\partial \mu}
{\partial n_{\rm tot}}{\displaystyle |_T}$
for $n_{tot}\to \infty$ equals the coupling constant
$g$ according to both sides of Eq. (\ref{csr2}). The border of
instability $\frac{\partial \mu}
{\partial n_{\rm tot}}{\displaystyle |_T}=0$
occurs where the density autocorrelation function
has its pole, i.e. for $2g\chi^0_{\rm nn}(0,0)=-1$.
It should be noted that this is the same point where the
longitudinal order-parameter response-function $G_+({\bf k},0)$ has its singularity, and also where $(\partial n_c/\partial\mu)_T$ becomes singular,
as follows from the relation $\partial\mu/\partial n_c=g(1+2g\chi^0_{nn}(0,0))$, which can be easily proven. The infinite slope $\frac{\partial \mu}
{\partial n_{\rm tot}}{\displaystyle |_T}=\infty$
of the unstable branch occurs where
$\chi_{\rm nn}(0,0)$ vanishes, i.e. for $g\chi^0_{\rm nn}(0,0)=-1$.

\section{Dynamics at long wavelengths}
\label{sec:numerics}

Now we want to investigate the dynamics at long wavelengths.
We consider the limit $\omega\to 0$ and $|{\bbox{k}}|\to 0$
and introduce the complex velocity of sound by
$\omega/|{\bbox{k}}|=c\,$.
The dispersion-relation
in the condensed phase, given by the poles of the
density autocorrelation function and the Green's function,
takes the form
\begin{equation}
    (c^2-c_B^2)
    = (c^2+c_B^2)\,2g
\,\chi_{\rm nn}^0({\bbox{k}},\omega) \,,
\label{sound.below}
\end{equation}
where we introduced the Bogoliubov speed of sound
$c_B=\sqrt{g\,n_c/m}\,.$
Eq. (\ref{sound.below}) will be the central equation we shall analyze in this
section.
The response function (\ref{eq:chi0}) for a homogeneous system
can be rewritten in the well-known form
\begin{equation}
   \chi_{\rm nn}^0({\bbox{k}},\omega)=
   \int {d^3p \over (2\pi)^3}\,
   {f_0(E^{HF}(\bbox{p})) -f_0(E^{HF}(\bbox{p}+{\bbox{k}})) \over
   \omega-(E^{HF}(\bbox{p}+{\bbox{k}}) -E^{HF}(\bbox{p}))
} \,,
\label{chi.hom}
\end{equation}
where $f_0(E^{HF})=(\exp(\beta (E^{HF}))-1)^{-1}$ is the Bose-distribution.
For $T\le T_c$ we have
$E^{HF}({\bbox{k}})=
{\bbox{k}}^2/2m+gn_c$
and $\bar\mu=-gn_c\,,$.

The dispersion-relation for the density-fluctuations
in the uncondensed phase is formally also contained
in Eq.(\ref{sound.below}) putting there $c_B=0$ and has one branch with $c=0$, corresponding to the fact that the single-particle dispersion-law in the uncondensed phase is proportional to $k^2$, and another branch given by
\begin{equation}
   1 = 2g
\,\chi_{\rm nn}^0({\bbox{k}},\omega) \,.
\label{sound.above}
\end{equation}
The response function (\ref{chi.hom}) in that region
is evaluated
with the effective fugacity
$z=e^{\beta\bar\mu}\,,$
according to
$\bar\mu=\mu-2g n_{\rm tot}=\mu^{(0)}\,,$
where the chemical potential $\mu$ is determined
thermodynamically
as described above.

For long wavelengths,
i.e. for wave vectors
with $|\bbox{k}|\lambda_{\rm th}\ll 1$,
we can approximate
in the denominator of the response function (\ref{chi.hom})
  $ E^{HF}({\bbox{p}}+
{\bbox{k}}) -E^{HF}({\bbox{p}})
   \approx 
\bbox{ k\cdot p}/m.$
The mean field interactions
for our two models
cancel in this difference.
The difference of the Bose factors in the numerator
can be be approximated by a gradient.
To evaluate
(\ref{chi.hom}) after these approximations further we choose
${\bbox{k}}$ in x-direction
and integrate first
in $p_y, p_z$
with the result
\begin{displaymath}
  \chi_{\rm nn}^0(k_x,\omega)
  = \frac{1}{(2\pi)^2}
    \int dp_x\,\,
    \frac{p_x}
    {\omega/|\bbox{k}| -p_x/m}\,f_0(E^{HF}(p_x)) \,.
\end{displaymath}
The response function only depends on the ratio
$\omega/|\bbox{k}|\,,$
which defines
the (generally complex) 'speed of sound'
defined by the ratio
$c=\omega/|\bbox{k}|$ in the long wavelength limit.
So equation (\ref{sound.below})
is an implicit equation for  $c\,.$
Scaling $p_x$ by $\sqrt{2mk_BT}$
and
measuring $c$
in units of the thermal velocity
$c_T=\sqrt{2k_BT/m}\,$,
the response function reads
\begin{equation}
  \chi_{\rm nn}^0(c)
  = \frac{
\lambda_{\rm th}^{-3}}{k_BT}\,\,\tilde\chi_{\rm nn}^0(c/c_T)
\label{chi.scal}
\end{equation}
with the dimensionless response function
\begin{equation}
  \tilde\chi_{\rm nn}^0(s)
  =\frac{1}{\sqrt{\pi}}
   \int\nolimits_{-\infty}^{\infty} dt\,\,
    \frac{t}
    {s -t}\,\frac{z}{e^{t^2}-z} \,,
\label{chi.dimlos}
\end{equation}
defined in the upper half of the complex $s$-plane, where the variable $s$
denotes the quotient $s=\omega/(c_T\,|\bbox{k}|)$.
The temperature dependence
resides in the prefactor in (\ref{chi.scal}) and in the fugacity $z$.

The integral can be evaluated
by the methods of residua, see \cite{Sz74}, and we refer to this paper for further details of the calculation. The result is
 the expression for the dimensionless response function
\begin{eqnarray}
  \tilde\chi_{\rm nn}^0(s)
  &=&-g_{1/2}(z)
   -\sqrt{\pi}i\,\frac{sz}{e^{s^2}-z}
   +\sqrt{\frac{\pi}{\gamma}}\,\,\frac{s^2}{s^2+\gamma}
    \nonumber\\
   &&+\sqrt{\pi}i\,
    \sum\nolimits_{n=1}^{\infty}\,
    \frac{s^2}{s^2-a_n^2}\,
    \frac{1}{a_n}
   +\frac{s^2}{s^2-b_n^2}\,
    \frac{1}{b_n} \,.
    \label{chi.resd}
\end{eqnarray}
with poles at
\begin{eqnarray}
\left.
\begin{array}{l}
a_n \\ b_n
\end{array}
\right\}
=
i\sqrt[4]{4\pi^2\,n^2+\gamma^2}\,\,e^{\pm i\phi_n/2}
\nonumber
\end{eqnarray}
and with $\gamma=-\beta\bar\mu=|\beta\bar\mu|=-\ln(z)\,,$
and
$\phi_n=\arctan(2\pi n/\gamma)$
for $n\ge 0$.
In the Hartree-model \cite{Sz74} $\gamma$ was $0$ in the condensed phase, while
here it is non-zero in the condensed and uncondensed phase. This may seem like a small difference, but his would be misleading, because a non-zero value of $\gamma$, even if it is small, leads to qualitative and by no means small differences for the response-function in the
limit $k\rightarrow 0$.
For small $\gamma$ one can expand
$
  g_{1/2}(e^{-\gamma}) = \sqrt{\pi/\gamma}+\zeta(1/2) +O(\gamma)\,.
$

When we ask for solutions of (\ref{sound.below},\ref{sound.above}),
we are looking for eigen-modes of the system,
which, for physical reasons, have to decay (rather than grow) exponentially
 and
the poles we look for have therefore to be located in the lower complex
half-plane.
In Eq. (\ref{sound.below})
the analytical continuation of the integral (\ref{chi.dimlos})
from the upper to the lower complex half-plane of $s$ has therefore
to be used.

\begin{figure}
\begin{center}
\epsfig{file=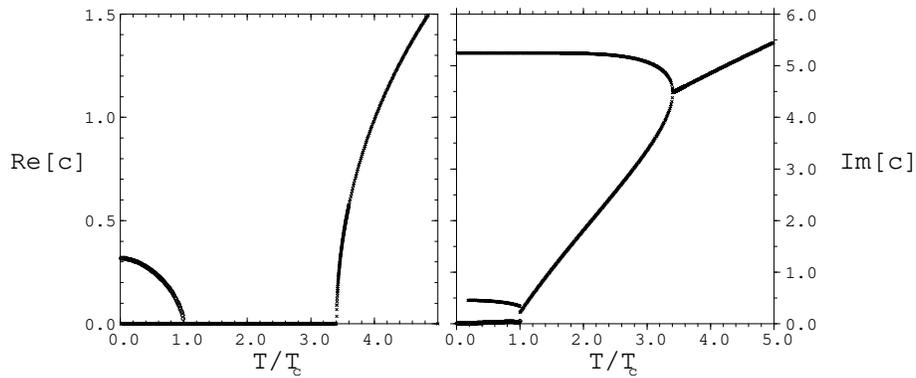, height=12cm, angle=270}
\end{center}
\caption{Real part (left) and imaginary part (right)
of the different branches of the velocity of sound $c=\omega/|\bbox{k}|$
from (\ref{sound.below}) and (\ref{sound.above})
in units $\sqrt{k_BT_c/m}$
for the interaction $gn_{\rm tot}/k_BT_c=0.1$
depending on the temperature $T/T_c\,.$
}
\label{v0=0.1.hf}
\end{figure}

With this explicit representation of the response function
we are now able to determine the solutions
of (\ref{sound.below}) and of (\ref{sound.above})
below and above the phase transition
numerically.
Since the terms in the sum
in (\ref{chi.resd}) decay only as $n^{-3/2}\,,$
the sum converges very slowly
and it is not possible
to cut the sum at some large finite $n_{\rm cut}\,.$
Instead we can use a continuum approximation
integrating over all terms in the range
$n \in [n_{\rm cut},\infty [\,.$
As a numerical check we have reproduced all our data
with the alternative representation of the response function
by performing integral (\ref{chi.dimlos}) numerically
with $s$ in the  lower complex half-plane
and subtracting the term for the analytical continuation:
\begin{equation}
  \tilde\chi_{\rm nn}^0(s)
  =\frac{1}{\pi^{1/2}}
   \int\nolimits_{-\infty}^{\infty} dt\,\,
    \frac{t}
    {s -t}\,\frac{z}{e^{t^2}-z}
   -2\sqrt{\pi} i\,\frac{sz}{e^{s^2}-z}
\,.
\label{chi.num}
\end{equation}
The numerical solutions are plotted in Figs.\ref{v0=0.1.hf},\ref{v0=0.3.hf}.
These figures show the real- and imaginary parts of $c$ for
various branches of damped modes as a function of temperature above and below
the phase-transition, and how these branches bifurcate as the poles
corresponding to these modes move in the complex plane.
For the weakest interaction strength, shown in
Fig.\ref{v0=0.1.hf}, various approximations are possible to
permit an analytical understanding of most of the structure shown. This will
be described in the following section.
Here we discuss the numerical results shown in the figures. We use the condensation temperature $k_BT_c$ of the ideal Bose-gas as a convenient energy-scale near which the phase-transition occurs.

Let us begin with  the case of weak interaction
with $gn_{\rm tot}/k_BT_c=0.1$
for our two models,
 shown
in  Fig.\ref{v0=0.1.hf}.
At high temperatures $T\gg T_c$
we typically get complex solutions for $c$
with $|c|\gg c_T\,,$
and with finite real and imaginary part.
Then, at a certain temperature $T_0$, which is above $T_c$ for very weak
interaction but moves below $T_c$ for stronger ones as seen in
Fig. \ref{v0=0.3.hf},
the real part of the velocity of sound vanishes,
the mode becomes over-damped below $T_c$,
and the imaginary part  bifurcates into two different branches,
as seen in Fig. \ref{v0=0.1.hf}.
Density-waves with a certain wave vector no longer propagate,
but decay as a mere relaxation with two different decay-rates describing
a short-time decay and a long-time decay.
Both damping-rates are proportional to the wave vector.
An analytical understanding of the bifurcation based on
a suitable approximation to the response function
for large arguments $|s|\gg 1$ will be provided in the next section.

The finite jump (\ref{ncond.disc}) of
the condensate density
at the first order transition in the model
results in discontinuities of the two purely imaginary thermal branches.
The discontinuity in the lower branch is visible in Fig. \ref{v0=0.1.hf},
while  in the upper branch it is very small and not discernible.
Even at lower temperatures this upper branch hardly deviates
from the result above $T_c$ as can be seen in  Fig. \ref{v0=0.1.hf}.
The reason for this is that
for weak interaction
(\ref{c.im.upper})
also implies
$|c|\gg c_B\,$,
and the dispersion-relation (\ref{sound.below}) below $T_c$
reduces to
the equation (\ref{sound.above}) above $T_c\,$.
So $c$ from (\ref{c.im.upper})
turns out to be also a solution
of (\ref{sound.below}) below $T_c$
and can be seen in  Fig. \ref{v0=0.1.hf}
as the solution independent of the temperature
for the whole condensed region.

Below the phase-transition a new propagating branch
with non-vanishing real part of $c$ appears, which is the Bogoliubov mode and,
as discussed in the next section, has a
velocity close to $c_B$, as is also visible in
Fig.\ref{v0=0.1.hf}.
In the units chosen in the figures
this Bogoliubov branch for
$T\to 0$ converges
to the value
$\sqrt{gn_{\rm tot}/k_BT_c}\,$.

Let us now turn to the case of slightly stronger
interaction in Fig. \ref{v0=0.3.hf}.
The most remarkable new feature
compared to the very weakly interacting case is,
that the bifurcation of the thermal branch
occurs now at some temperature $T_0$ below the phase transition.
We can see the real parts of the propagating
thermal branch and the Bogoliubov branch crossing
at some temperature between $T_0$ and $T_c$.
Though the mode described by the thermal branch is still propagating
it is strongly damped,
its real part being smaller than its imaginary part.
The bifurcation of the Bogoliubov branch near $T_c$
does not occur anymore,
the Bogoliubov sound remains propagating
at the phase transition.
Again the smaller of the two damping rates which have bifurcated from the
thermal branch
is still larger than the damping of the Bogoliubov sound.
However, the discontinuity in the condensate density
at the phase transition
already amounts to $40\%$ of the total density,
so this behavior near $T_c$ has only a
restricted physical meaning.

\section{Analytical solution of the dispersion relation for weak
interaction}\label{analytic}

\subsection{Thermal branch}

The bifurcation shown in Fig.\ref{v0=0.1.hf}
can be understood
by the following approximation to the response function:
for large arguments $|s|\gg 1$ we can expand
the integrand in (\ref{chi.num})
in powers of $(t/s)^2$
and
get as the dominant behavior for large $|s|$
\begin{equation}
  \tilde\chi_{\rm nn}^0(s)
  = \frac{g_{3/2}(z)}{2 s^2}
   -2\sqrt{\pi} i\,\frac{s\,z}{e^{s^2}-z} +O\Big(\frac{1}{|s|^4}\Big)
\,.
\label{chi.ssgg}
\end{equation}
For weak interaction
$g\,n_{\rm tot}\ll k_BT$
and for $T>T_c$
the real part of $s$ is much smaller than the imaginary part,
as seen numerically,
and the full response function can be approximated
by the second term in (\ref{chi.ssgg}) only.
This approximate response function, taken as a function
of a purely imaginary argument,
is a real convex function,
and only if condition (\ref{sound.above})
is fulfilled at its minimum,
a purely imaginary solution is possible.
From the temperature dependence of the chemical potential above $T_c$
it follows,
that this is possible only below some temperature $T_0,$
which one can determine as $T_0/T_c=3.37$
for $g\,n_{\rm tot}/k_BT_c=0.1\,,$
for example.
Above $T_0$ the approximate response function
gives rise to
the characteristic growth of the real part of $c$,
see Fig. \ref{v0=0.1.hf}.
In the limit $|s|\gg 1$
we even can approximate
$
  \tilde\chi_{\rm nn}^0(s)
  \approx 2\sqrt{\pi} i\,s\,,
$
and the simple solution
independent of the temperature follows as
\begin{equation}
  c = -\frac{i}{4\sqrt{\pi}}\,\frac{k_BT}{g\lambda_{\rm th}^{-3}}\,c_T
    = -\frac{i\pi
}{m^2g} \,.
\label{c.im.upper}
\end{equation}
This can be seen as the upper branch
of the bifurcated imaginary part
for small temperatures
in Fig. \ref{v0=0.1.hf}.
The lower branch decreases linearly with temperature
and finally enters the opposite region with $|s|=|c/c_T|\ll 1\,.$.
In this region
the behavior can be described by a different approximation
of the response function
\cite{Sz74}.
The first three leading terms of (\ref{chi.resd})
in $s,\gamma$
ordered by magnitude
can be summarized as
\begin{equation}
  \tilde\chi_{\rm nn}^0(s)
  = \frac{-i\sqrt{\pi}}{s+i\sqrt{\gamma}} -\zeta(\frac{1}{2})
   +\frac{i\sqrt{\pi}s}{2}
   + O(|s|^2)\,.
    \label{chi.ssll}
\end{equation}

From the dominant, first two terms
the approximate solution
in the limiting case $|c|\ll c_T$
above the phase transition follows as
\begin{equation}
  c = -i\,\frac{
      \sqrt{\frac{-2\mu}{m}}
    + g\lambda_{\rm th}^{-3}\,\sqrt{\frac{8\pi}{mk_BT}}}
     {1+ \frac{2g\lambda_{\rm th}^{-3}}{k_BT} \zeta(\frac{1}{2})} \,.
\label{c.im.lower}
\end{equation}
Except at high temperatures this expression agrees very well with the numerically determined data
in Figs.\ref{v0=0.1.hf}.
The behavior linear in $T-T_c$
is given by the first term in the numerator,
since the chemical potential can be expanded as
$ -\mu = \zeta(3/2)^2 \,9 k_B(T-T_c)^2 /16\pi T_c + O((T-T_c)^3)$,
the offset
is given by the second term in the numerator.

Finally we note, that the bifurcations discussed above can also be followed 
analytically if use is made of the approximation (\ref{chi.ssll}) for the density response function.

\subsection{Bogoliubov-branch}

Let us turn to the region below $T_c$ and show that
in both of our models we have a branch with $c$ very close to $c_B$.
In the  limit
$|c|\ll c_T$
we can use again the approximate response function
(\ref{chi.ssll}).
According to the Bogoliubov theory
at vanishing temperature
sound propagates with the Bogoliubov speed $c_B$.
Using this in (\ref{chi.ssll})
we see, that
$g\chi_{\rm nn}^0(c_B)$ is of the order $O(s_B,\gamma^{1/2})\ll 1$
with $s_B=c_B/c_T=\sqrt{\gamma/2}$.
So with $g\chi_{\rm nn}^0(c_B)\ll 1$
it follows
from (\ref{sound.below}),
that one solution is near the
Bogoliubov speed of sound
$c_B\,$.
The corresponding numerically determined solutions
can be seen in  Fig.\ref{v0=0.1.hf}

Being close to $c_B$ the Bogoliubov-branch
can be determined analytically from perturbation theory.
To this purpose we consider
the right hand side
in (\ref{sound.below})
with the response function (\ref{chi.ssll})
as a perturbation to
the Bogoliubov speed of sound
$c=c_B\,$.
We get
the complex correction
\begin{equation}
\label{hom.c.perturb}
    c-c_B
    = 2c_B\,\frac{g\lambda_{\rm th}^{-3}}{k_BT}\,
    \Big(\frac{-\sqrt{\pi}i}{s_B+i\sqrt{\gamma}}
    -\zeta(\frac{1}{2})
    +\frac{i\sqrt{\pi}s_B}{2}\Big) \,.
\end{equation}
The first term gives the leading order in $g$
and is proportional to the temperature.
The damping rate
of Bogoliubov sound waves with wave vector $q$
due to this term is
\begin{equation}
\label{dampf.hf}
    \Gamma_1 =
    \frac{4}{3} \,
    k_BTa
\,q \,,
\end{equation}
with the $s$-wave scattering length $a$ from $g=4\pi
a/m\,.$
This damping rate is of the same order as the damping
previously determined
in \cite{Sz74},
which only differs in the numerical prefactor
of one instead of $4/3\,,$
and also agrees approximately with the damping rate
in the Beliaev approximation extended to finite temperatures
by {\sl Shi and Griffin} \cite{ShiGr}, but obtained also by
other methods  in the intermediate
temperature  region  \cite{Liu,SP,SF,G2}.  The difference is that in
(\ref{dampf.hf}) the prefactor $4/3$ should be replaced by $3\pi/8$.
For the frequency shift we obtain
$\Delta\omega=-\sqrt{2}\Gamma$, a result which is in good agreement with
a result of Fedichev and Shlyapnikov \cite{SF}, who obtain $\Delta\omega/\Gamma=-(28/3\pi^{3/2})=1.67...$, and also of Giorgini \cite{G},
who gets $\Delta\omega/\Gamma\approx -1.8$
This shift, which is negative and linear in the temperature,
reduces the speed of sound
compared to the Bogoliubov approximation.
These results are in agreement with the measurements
of temperature dependent frequency shifts
of discrete modes \cite{Jin97},
which were found to have negative sign for
the $m=2$ mode,
and also for the $m=0$ mode at intermediate temperatures.
Using the full response function
and taking the inhomogeneity of the system into account
an explanation of these frequency shifts and the damping rates
was already given in \cite{Rd99b}.

\section{Discussion and Conclusion}
\label{conclusion}

Let us briefly summarize the results of this paper and then draw some further conclusions. We have presented here within the framework of the dielectric formalism a consistent microscopic model of the weakly interacting Bose-gas including exchange.
We have shown that a consistent treatment of 
exchange processes is achieved by using Hartree-Fock propagators for the internal lines of diagrams and summing up the {\it same} classes of diagrams
for {\it different} quantities. As far as the density correlation function is concerned we obtain results which, on the general level, are equivalent
to earlier reults by Minguzzi and Tosi \cite{Min}.
This agreement is nontrivial, because our starting point is quite different
from their's. However our treatment is more general than that in \cite{Min}
because it also gives the single-particle Green's functions, which by construction have the same poles as the density-correlation function, displaying a gapless single-particle spectrum, i.e. satisfying the Hugenholtz-Pines theorem.
The agreement of our result for the density correlation with that in \cite{Min} uncovers the diagrammatical basis of the 
equations written down there and therefore
opens up the possibility for future systematic
improvements. The rational basis of our 'model' or approximation is
its consistency with general requirements, which are
very nontrivial to be satisfied simultaneously. Thus we demonstrated explicitely that the compressibility sum-rule is satisfied, ensuring the consistency between statics and dynamics of the model; and the Ward-identities, were checked, ensuring the consistency between particle number
conservation (and the f sum-rule) and the spontaneously broken gauge-symmetry.

To these consistency checks, which were in detail discussed, a further one may be added which we have not yet discussed, but which we deem to be of no less importance, because
it derives from a further symmetry of the system - Galilei-invariance.
Galilei-invariance is most easily considered in a spatially confined system,
because it then simply implies the free motion of the center of mass,
if the confined system as a whole moves. In Bose-condensed systems spatial confinement is naturally achieved by imposing an external, but spatially fixed trapping potential. The system can then not move as a whole, but can still move
in the external potential. The motion of the center of mass is then no longer free, and in general it is not even separable from the other degrees of freedom
in quantum mechanics. However in the special case of an external harmonic potential the center of mass motion is separable and is simply a harmonic oscillation in the external potential. This fact is the content of the Kohn-theorem \cite{K}. In the limit where the spring-constants of the external harmonic potential are sent to zero the harmonic oscillation of the center of mass tends to the free motion required by Galilei-invariance. Therefore,
if the system satisfies the Kohn-theorem in a fixed external harmonic potential
(possibly with infinitesimal spring-constants), Galilei-invariance is ensured
if the external potential is switched off.
We wish to point out here, that in addititon to the other consistency-checks
already discussed, also this check is passed by the approximate model
discussed in the present paper. This has been shown in \cite{Kohn},
where the fulfillment of the Kohn-theorem within the approximate model
was explicitely demonstrated. It can be seen from the proof given there
that this test
is quite sensitive and would e.g. fail if. another definition of $n_{\rm th}$ would be chosen, e.g. by using in the definition the full Green's function rather than the Hartree-Fock propagators, or if propagators different from the Hartree-Fock propagators would be used in the internal lines.

Let us briefly discuss also  the limits of validity of our
treatment. The theory we have given is a mean-field theory and
therefore restricted to the collisionless domain $\omega\tau_{coll}\gg1$, where $\tau_{coll}\gg1$ is the mean collisiontime. It is also restricted to sufficiently high temperature $k_BT\gtrsim gn_c$, because in the low-temperature
domain $k_BT\le gn_c$ scattering processes at wave-number smaller than the inverse Bogoliubov coherence length $\xi_B=(8\pi n_ca)^{-1/2}$ make 
an important contribution,
and our use of the Hartree-Fock propagator for the internal lines
would lead to a qualitatively wrong temperature-dependence. The mean-field character of our theory
also prohibits its application close to the phase transition, which
occurs at a temperature close to $T_c$ for the weakly interacting Bose-gas.
The Ginzburg-criterion  for the validity of a mean-field description
here takes the form \cite{SF} $|T-T_c|/T_c\gtrsim(n_{tot}a^3)^{1/3}$. Indeed, our model,
if extrapolated to temperatures near $T_c$, would predict a first-order transition within the transition-region specified by this criterion, but since this is clearly outside the limit of validity, it is of course 
not a prediction of the model.
For some purposes, like following the fate of the various excitation branches
as the phase-transition temperature is crossed, it would certainly 
be nice to have also a mean-field model including exchange and satisfying all the consistency checks we have discussed {\it and} also giving a second-order
mean-field transition at a critical temperature near that of the ideal
Bose-gas. This goal, however, is not met by the
approximation we have discussed here, and further work may be required to
achieve it eventually.

In summary, the results obtained here identify the model we introduce as a rather satisfactory while still manageable microscopic description of a weakly interacting Bose-gas in the collisionless regime, except at very low temperatures and very near to $T_c$.

\begin{figure}
\begin{center}
\epsfig{file=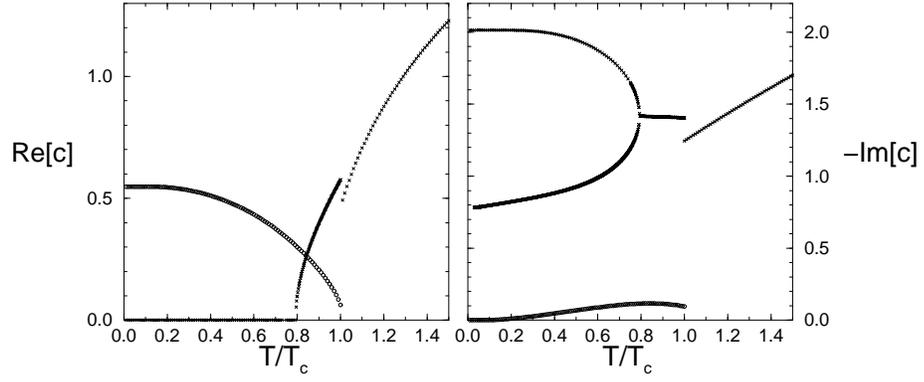, height=12cm, angle=270}
\end{center}
\caption{Real part and imaginary part
of the different branches of the velocity of sound $c$
in units $\sqrt{k_BT_c/m}$
for the interaction $gn_{\rm tot}/k_BT_c=0.3\,.$
}
\label{v0=0.3.hf}
\end{figure}

Besides checking in detail the consistency of our approximations, we have presented and discussed a detailed numerical and partially also
analytical study of the dispersion relation of the {\it joint} single-particle and density fluctuation
modes below and the {\it separate} single-particle and density fluctuation
modes above $T_c$. The results for the complex ratio $c=\omega/|k|$
have been summarized in Figs.\ref{v0=0.1.hf},\ref{v0=0.3.hf}. The dispersion relation found depends
qualitatively on the strength of the coupling. If the latter is weak like in
Fig.\ref{v0=0.1.hf}, then purely damped modes exist from a region above $T_c$
down to the low-temperature regime, besides the propagating and
weakly damped Bogoliubov-mode
which exists only below $T_c$. For stronger coupling, like in Fig.\ref{v0=0.3.hf},
there is a propagating, damped mode from above $T_c$ down to a finite temperature $T_0$ somewhat below $T_c$. Only below $T_0$ this mode
 also becomes purely overdamped, like for the weak-coupling case. The Bogoliubov mode in the
condensed phase exists also for strong coupling, only with higher frequency and larger damping.
 
Though the main purpose of the present work has been a theory for the homogeneous Bose gas it is interesting to contrast the results presented here with the measurements of and theoretical results for
the temperature dependence of discrete frequencies
for trapped condensates \cite{Jin97},\cite{Bij},\cite{S2}.
The results for the real part of the velocity of sound
in  Fig.\ref{v0=0.3.hf}
are  similar to
measurements of the frequency shifts and damping-rates for the $m=0$ and $m=2$ modes at intermediate temperatures \cite{Jin97}.
The measured frequency of the $m=0$ mode was found to first decrease
with increasing temperatures for low temperatures,
but then it suddenly increased again
at higher temperatures
$T\gtrsim 0.6\,T_{\rm crit}\,.$
It was suggested,
that the increase might be due to the crossing
of the Bogoliubov mode with another mode of the thermal cloud \cite{Jin97},\cite{Bij},\cite{S2}.
Such a second mode was actually found
in \cite{Bij},\cite{S2}
using a  solution
of the kinetic equations.
Here we also found a second branch
of the velocity of sound for the homogeneous system,
see  Fig.\ref{v0=0.3.hf},
but it is found to be strongly damped, in qualitative distinction
from what is found in the trapped system.
However, a measurement along the lines of \cite{Andrews}, testing the local properties of the Bose-gas, could actually check our results for the
thermal branch of the collective modes in the homogeneous system.
In fact it was already remarked in \cite{Andrews} that above $T_c$ there was no clear evidence for a propagating sound-wave, which is in qualitative agreement
with the non-propagating nature of the collective mode above $T_c$ which
we have found in the present paper. 

\section*{acknowledgments}
The research results were attained with the assistance of the Humboldt
Research Award to one of us (P.Sz.) and through support by a project of the
Deutsche Forschungsgemeinschaft and the Hungarian Academy of Sciences under
grant No 130. Support by the Deutsche Forschungsgemeinschaft
within its Sonderforschungsbereich 237 'Unordnung und grosse Fluktuationen',
and by the Hungarian National Research Foundation under grant No OTKA T029552
is also gratefully acknowledged.

\appendix
\section*{}
\label{app:A}

\begin{figure}
\begin{center}
\epsfig{file=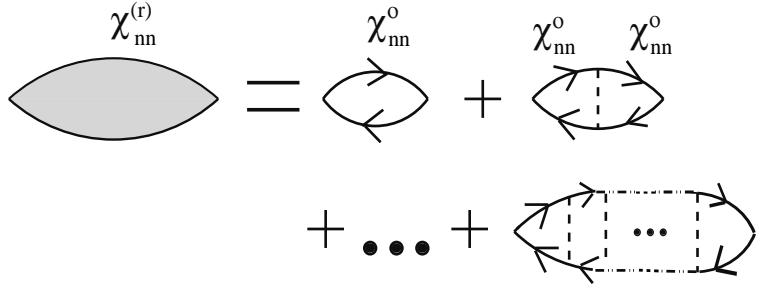, height=10cm, angle=90}
\end{center}
\caption{Diagrams contributing to   $\chi^{(r)}_{\rm nn}$.}
\label{fig:chila}
\end{figure}

\begin{figure}
\begin{center}
\epsfig{file=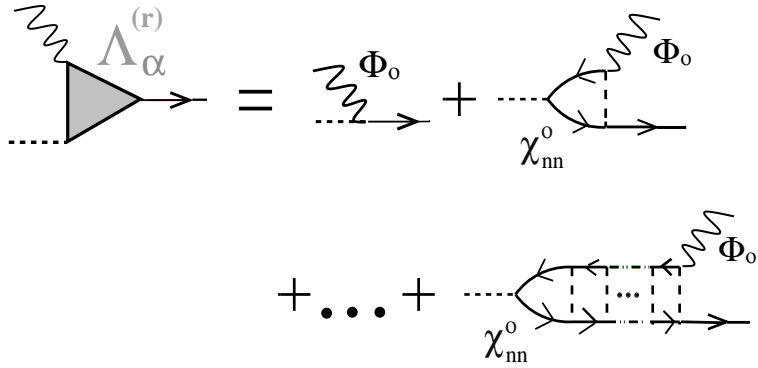, height=10cm, angle=90}
\end{center}
\caption{Diagrams contributing to  $\Lambda^{(r)}_\alpha$.}
\label{fig:vertexla}
\end{figure}

\begin{figure}
\begin{center}
\epsfig{file=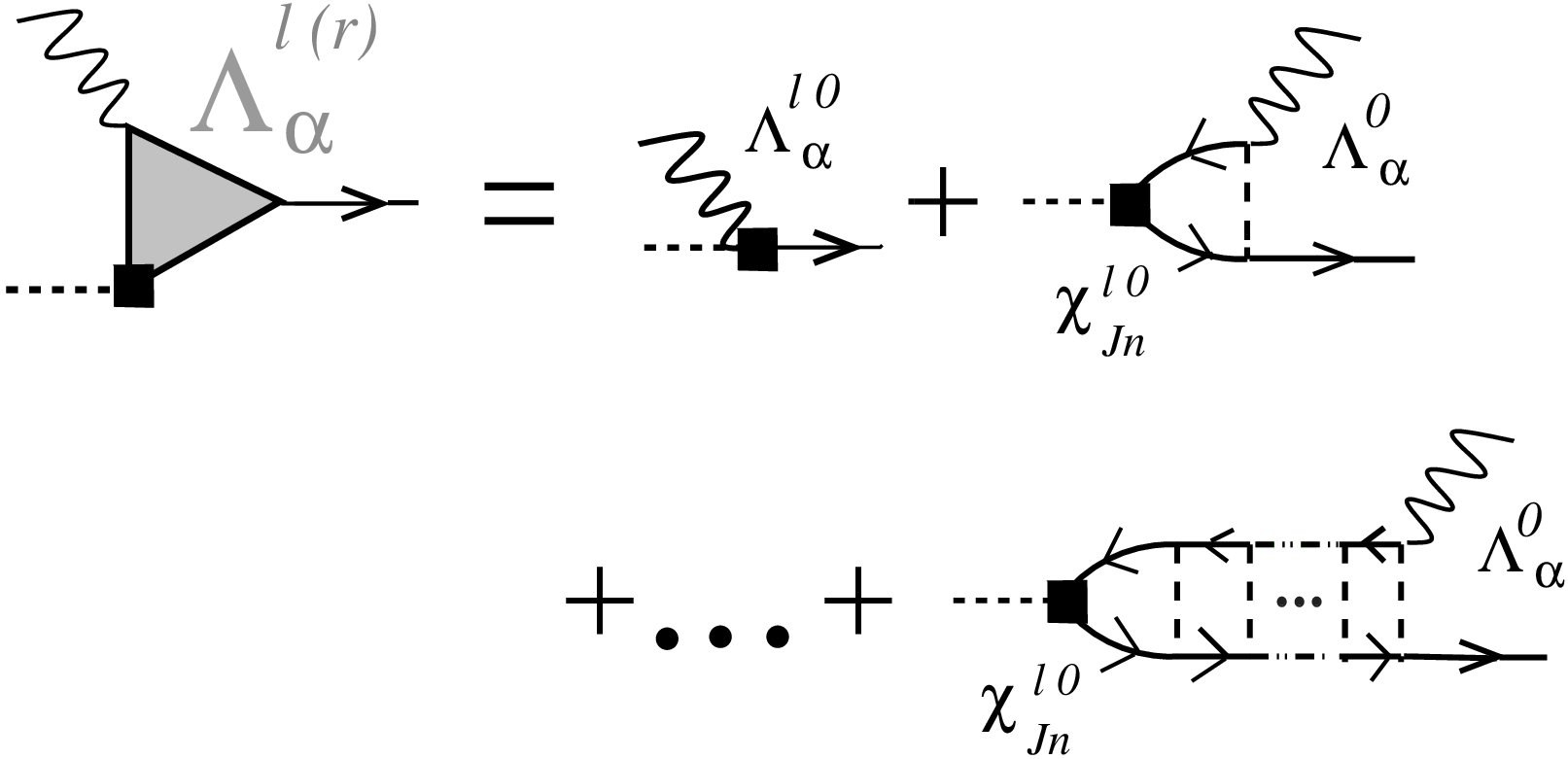, height=10cm, angle=90}
\end{center}
\caption{Diagrams contributing to $\Lambda^{l (r)}_\alpha$.\\
The symbol $\blacksquare$ denotes the longitudinal component of the gradient.}
\label{fig:curvertla}
\end{figure}

\begin{figure}
\begin{center}
\epsfig{file=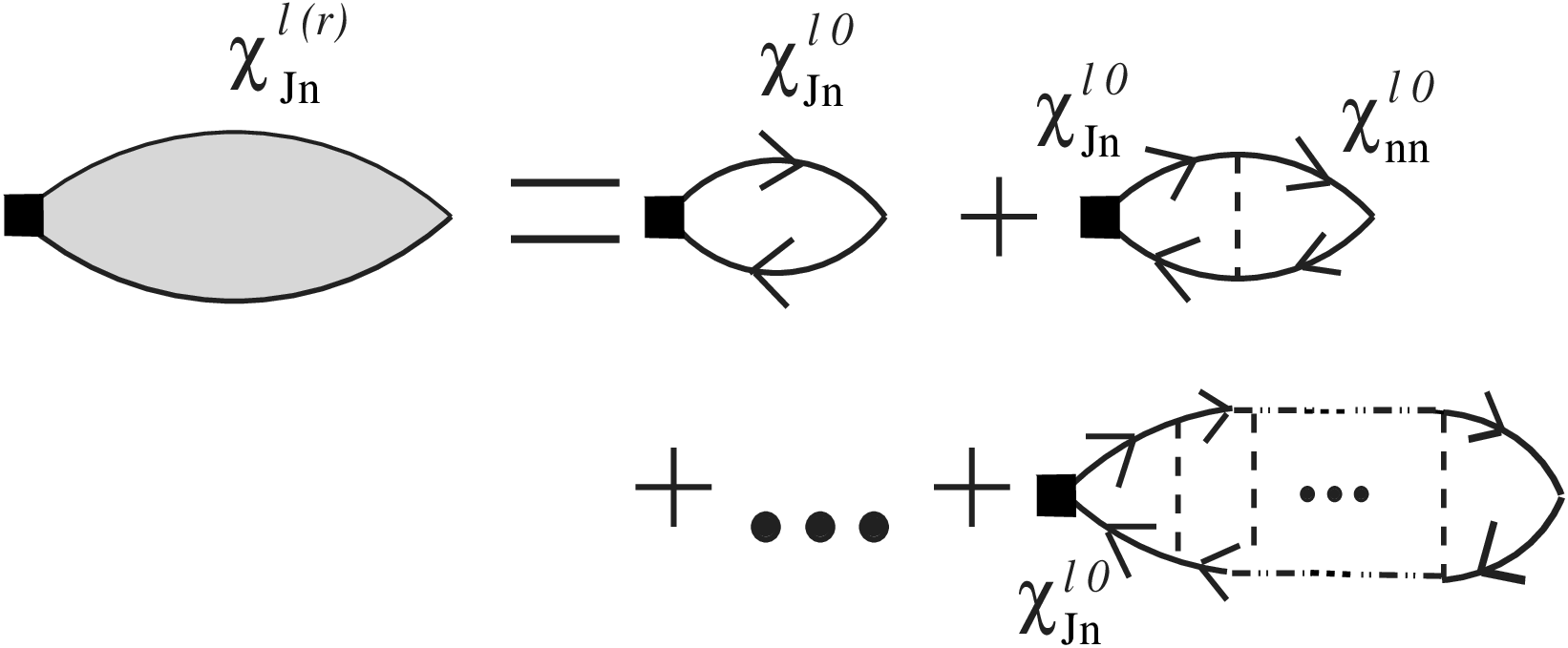, height=10cm, angle=90}
\end{center}
\caption{Regular current-density autocorrelation function
$\chi^{l\,(r)}_{Jn}$ corresponding to $\chi^{(r)}_{\rm nn}$ }
\label{fig:curchila}
\end{figure}

We wish to show here
that the identities (\ref{ward1}-\ref{ward3}) are indeed fulfilled in our model approximations. The specific choice of the building blocks for the
vertex functions
${\Lambda}^{(r)}_\alpha$ and ${\Lambda}^{l\,(r)}_\alpha$
simplifies the structure of   these identities considerably.
First the second term on the right hand side of the second Ward identity
vanishes due to the agreement of
${\Lambda}^{(r)}_1$ and ${\Lambda}^{(r)}_2$.
Second, rewriting ${\Lambda}^{(r)}_\alpha$ and
${\Lambda}^{l\,(r)}_\alpha$ in the forms:
\begin{eqnarray}
\label{tlam}
{\Lambda}^{(r)}_\alpha=\Lambda^0_\alpha+\Lambda^{1}_\alpha&=&
\sqrt{n_c}+g
{\chi}^{(r)}_{\rm nn}\sqrt{n_c}\\
\label{tlaml}
{\Lambda}^{l\,(r)}_\alpha=\Lambda^{l\,0}_\alpha
+{\Lambda}^{l\,1}_\alpha&=&
\alpha
\frac{|\bbox{q}|}{2}\sqrt{n_c}+g
{\chi}^{l\,(r)}_{Jn}\sqrt{n_c}
\end{eqnarray}
(compare Fig.\ref{fig:vertexla} and Fig.\ref{fig:curvertla})
we obtain  $
\sqrt{n_c}\,[{\Lambda}^{l\,(r)}_1-{\Lambda}^{l\,(r)}_2]=
\sqrt{n_c}\,[{\Lambda}^{l\,0}_1-{\Lambda}^{l\,0}_2]=
|\bbox{q}|n_c$
and the third Ward identity reduces to
\begin{equation}
\label{ward32}
\omega
{\chi}^{l\,(r)}_{Jn}(\bbox{q},\omega)
=\frac{|\bbox{q}|}{m}
[{\chi}^{l\,(r)}_{JJ}(\bbox{q},\omega)+
m\, n_{\rm th}].
\end{equation}
Furthermore,
in the first Ward identity the terms $\sqrt{n_c}\,
\omega$ and
$(-1
)\sqrt{n_c}\,\alpha\,
\bbox{q}^2/2m$ appearing in the bracket on the right hand side are
cancelled by the contributions $
\omega{\Lambda}^0_\alpha$ and
$|\bbox{q}|/m\,{\Lambda}^{l\,0}_\alpha$ respectively.
Since the contributions of ${\Sigma}^{1\,(r)}_{\alpha\beta}$ are
independent of $\alpha$ and $\beta$ the
difference ${\Sigma}^{1\,(r)}_{1\alpha}
-{\Sigma}^{1\,(r)}_{2\alpha}$ vanishes.
Additionally we note  that
the differences of  the Gross-Pitaevskii self-energies are  given by
${\Sigma}^0_{1\alpha}
-{\Sigma}^0_{2\alpha}=\alpha\mu$ resulting in a simplified
expression of the first Ward identity
\begin{equation}
\label{ward12}
\omega{\Lambda}^{1}_\alpha(\bbox{q},\omega)=\frac{|\bbox{q}|}{m}
{\Lambda}^{l\,1}_\alpha(\bbox{q},\omega).
\end{equation}
Recalling the decomposition in Eqs. (\ref{tlam},\ref{tlaml})
and dividing ${\Lambda}^{1}_\alpha$ and
${\Lambda}^{l\,1}_\alpha$ by their common factor $g
\sqrt{n_c}$
the  proof of  the first and second Ward identities is reduced
to  the check of the relation
\begin{equation}
\omega{\chi}^{(r)}_{\rm nn}=\frac{|\bbox{q}|}{m}{\chi}^{l\,(r)}_{Jn}.
\end{equation}
Using the decompositions (Fig. \ref{fig:chila} and \ref{fig:curchila})
\begin{eqnarray}
{\chi}^{(r)}_{\rm nn}={\chi}^0_{\rm nn}
\left(1+g
{\chi}^{(r)}_{\rm nn}\right)\\
\label{chilrdec}
{\chi}^{l\,(r)}_{Jn}={\chi}^{l\,0}_{Jn}
\left(1+g
{\chi}^{(r)}_{\rm nn}\right)
\end{eqnarray}
we just have to demonstrate the equivalent relation $
\omega{\chi}^0_{\rm nn}-
(|\bbox{q}|/m) \,{\chi}^{l\,0}_{Jn}=0$,
which can be done in a similar way as in \cite{Griffin} (see also
\cite{TG}) by replacing the free-particle Green's functions used there by the Hartree-Fock Green's functions $G^{HF}$ of eq.(\ref{GHF}).
In completing this proof  we only need an energy dispersion law
of the form $\varepsilon({\bbox{p}})\sim \left[\bbox{p}^2/2m+const.\right]$
valid for  the free-particle Green's function and the Hartree-Fock
Green's function used by us.

\begin{figure}
\begin{center}
\epsfig{file=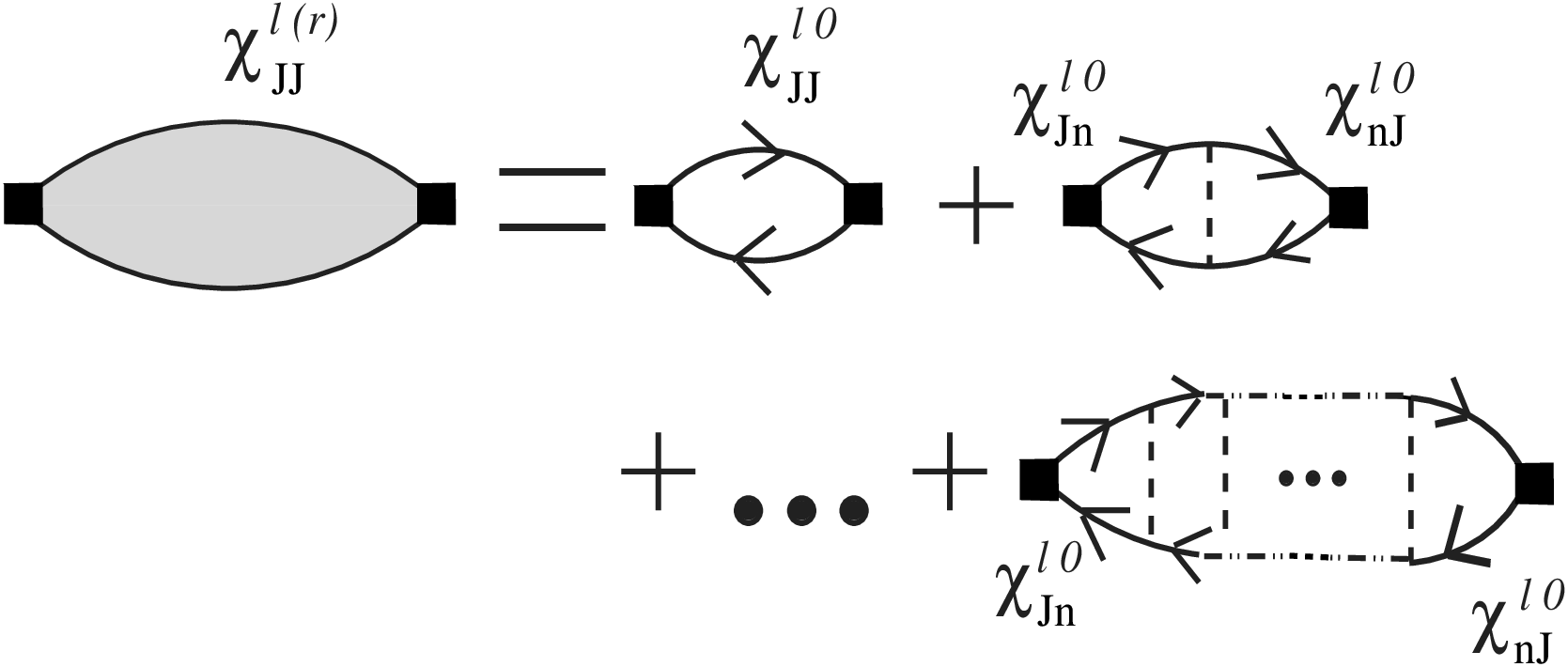, height=10cm, angle=90}
\end{center}
\caption{Regular current-current autocorrelation function
$\chi^{l\,(r)}_{JJ}$ corresponding to $\chi^{(r)}_{\rm nn}$ }
\label{fig:curcurchila}
\end{figure}

The third Ward identity can be simplified by the decompositions
(Fig. \ref{fig:curchila} and \ref{fig:curcurchila})
\begin{eqnarray}
{\chi}^{l\,(r)}_{Jn}&=&{\chi}^{l\,0}_{Jn}+{\chi}^{l\,0}_{Jn}
\frac{g
}{(1-g
{\chi}^{0}_{\rm nn})}{\chi}^{l\,0}_{\rm nn}\\
{\chi}^{l\,(r)}_{JJ}&=&{\chi}^{l\,0}_{JJ}+{\chi}^{l\,0}_{Jn}
\frac{g
}{(1-g
{\chi}^{0}_{\rm nn})}
{\chi}^{l\,0}_{nJ}
\end{eqnarray}
which reduces its proof  to the
check of the condition
\begin{equation}
\label{ward33}
\omega{\chi}^{l\,0}_{Jn}(\bbox{q},\omega)-\frac{|\bbox{q}|}{m}
{\chi}^{l\,0}_{JJ}(\bbox{q},\omega)=
|\bbox{q}|\, n_{\rm th}.
\end{equation}
After multiplication with $|\bbox{q}|/m$ we obtain:
\begin{eqnarray}
\nonumber
&&
\omega\frac{|\bbox{q}|}{m}{\chi}^{l\,0}_{Jn}(\bbox{q},\omega)-\frac{\bbox{q}^2}{m^2}
{\chi}^{l\,0}_{JJ}(\bbox{q},\omega)\\ \nonumber&
=&\int\frac{d{\bbox{p}}^3}{(2\pi
)^3}
\left(
\omega-\frac{\bbox{q}}{m}({\bbox{p}}+
\frac{\bbox{q}}{2})\right)\left[\frac{f_0(\varepsilon({\bbox{p}}))
-f_0(\varepsilon
({\bbox{p}}+\bbox{q}))}{\omega-\frac{\bbox{q}}{
m}({\bbox{p}}
+\frac{\bbox{q}}{2})}\right]\frac{\bbox{q}}{m}({\bbox{p}}+
\frac{\bbox{q}}{2})\\ \nonumber&
=&\int\frac{d{\bbox{p}}^3}{(2\pi
)^3}\frac{\bbox{q}}{m}({\bbox{p}}+
\frac{\bbox{q}}{2})\left[f_0(\varepsilon({\bbox{p}}))
-f_0(\varepsilon
({\bbox{p}}+\bbox{q}))\right]
=\frac{
\bbox{q}^2}{m}n_{\rm th},
\end{eqnarray}
which completes the proof.




\end{document}